\documentclass[12pt,preprint]{aastex}
\begin{document}

\title{Quasars and the Hubble Relation}

\author{H. Arp}
\affil{Max-Planck-Institut f\"ur Astrophysik, Karl Schwarzschild-Str.1,
  Postfach 1317, D-85741 Garching, Germany}
 \email{arp@mpa-garching.mpg.de}

\begin{abstract}

If active galaxies are defined as extragalactic objects with appreciably non
thermal spectra then a continuity exists in redshift from the highest
redshift quasars to low redshift Seyferts, AGNs and allied galaxies. 

Evidence is discussed for this sequence to be an evolutionary track
with objects evolving from high to low intrinsic redshift with time. 
At the end of this evolution the objects are nearly the same age as
our own galaxy and they come to rest on the traditional Hubble
relation.                                                                     
                                                                              
\end{abstract}

{\bf Introduction}

In 1963 some blue, stellar appearing objects in the sky were found to
have high redshifts. They had been observed initially because they
were radio sources, but then radio quiet and finally X-ray, stellar
appearing sources were found with even higher redshifts. They were
called Quasi Stellar Objects, soon shortened to QSOs or quasars. The
main interest lay in their high redshifts which were interpreted as
requiring great distances and therefore much higher luminosities than
previously derived for any extragalactic objects. 

Spurred on by the promise of learning about far reaches of
the Universe, researchers of that day competed in discovering and
analyzing more of these high redshift quasars. A disappointing result,
however, soon became evident. These supposedly most distant objects
showed no redshift - apparent magnitude (Hubble) relation. In the z - m
diagram the clump of quasar points showed too much dispersion in
redshift over a small range in apparent magnitude.
  
But in all the attention given to the high redshift objects little
notice had been taken of objects that looked like QSOs except they had
intermediate to low redshifts. They had been given various names such
as N galaxies, compact galaxies, compact and radio nuclei, emission
line and/or Seyfert spectra. All of them shared properties with
quasars and formed a link between quasars and nearby galaxies. When
these were plotted in the z - m diagram it became apparent that there
was a broad relation between redshift and apparent magnitude. 

This relation can be seen here in Fig. 1, the diagram of QSO and
QSO-like objects which were known at that time (Arp 1968). Most
recently Morley Bell (2007) has plotted all currently known quasars
and active galaxies (106,958) in the diagram which is shown here in
Fig. 2. This important result shows the large number of current
points confirms the earlier seen sharp rise in redshifts between apparent
magnitude 16 and 20. The problem is still, however, that the relation
is broader and overall shows a different slope than the standard
Hubble Relation. 

\begin{figure}[ht]
\includegraphics[width=11.0cm]{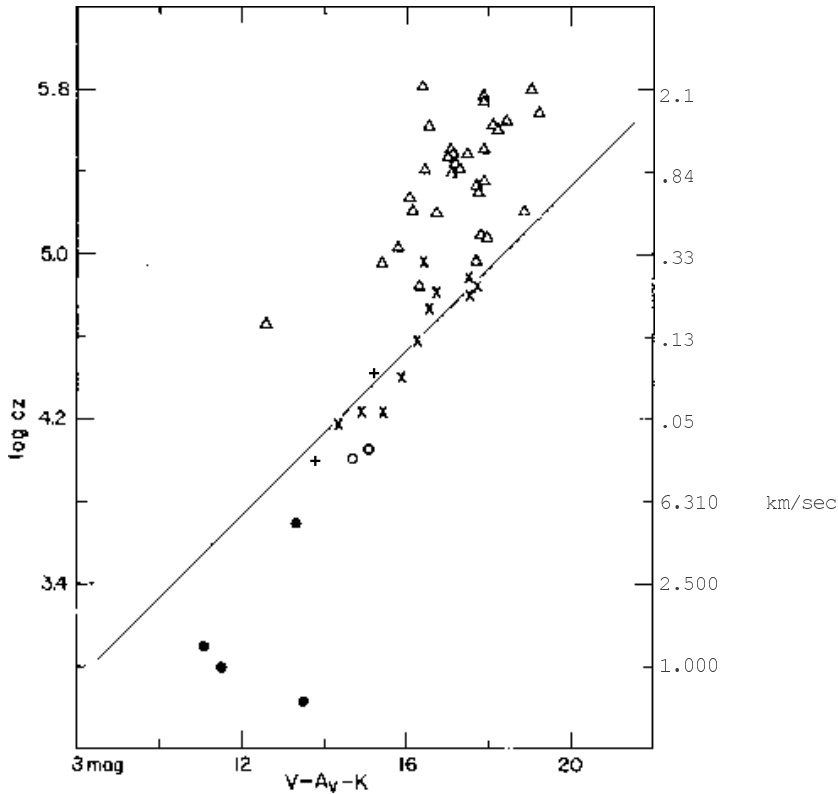}
\caption{Redshift-apparent magnitude diagram for QSOs ({\it
triangles}) compact Seyfert spectra ({\it plus signs}), radio
galaxies with compact nuclei ({\it crosses}), two Zwicky compact
galaxies ({\it open circles}) and Seyfert galaxy nuclei ({\it filled
circles}). The Hubble relation line is for radio E galaxies. (Diagram
from Arp 1968).
\label{fig1}}
\end{figure}

\begin{figure}[ht]
\includegraphics[width=10.0cm]{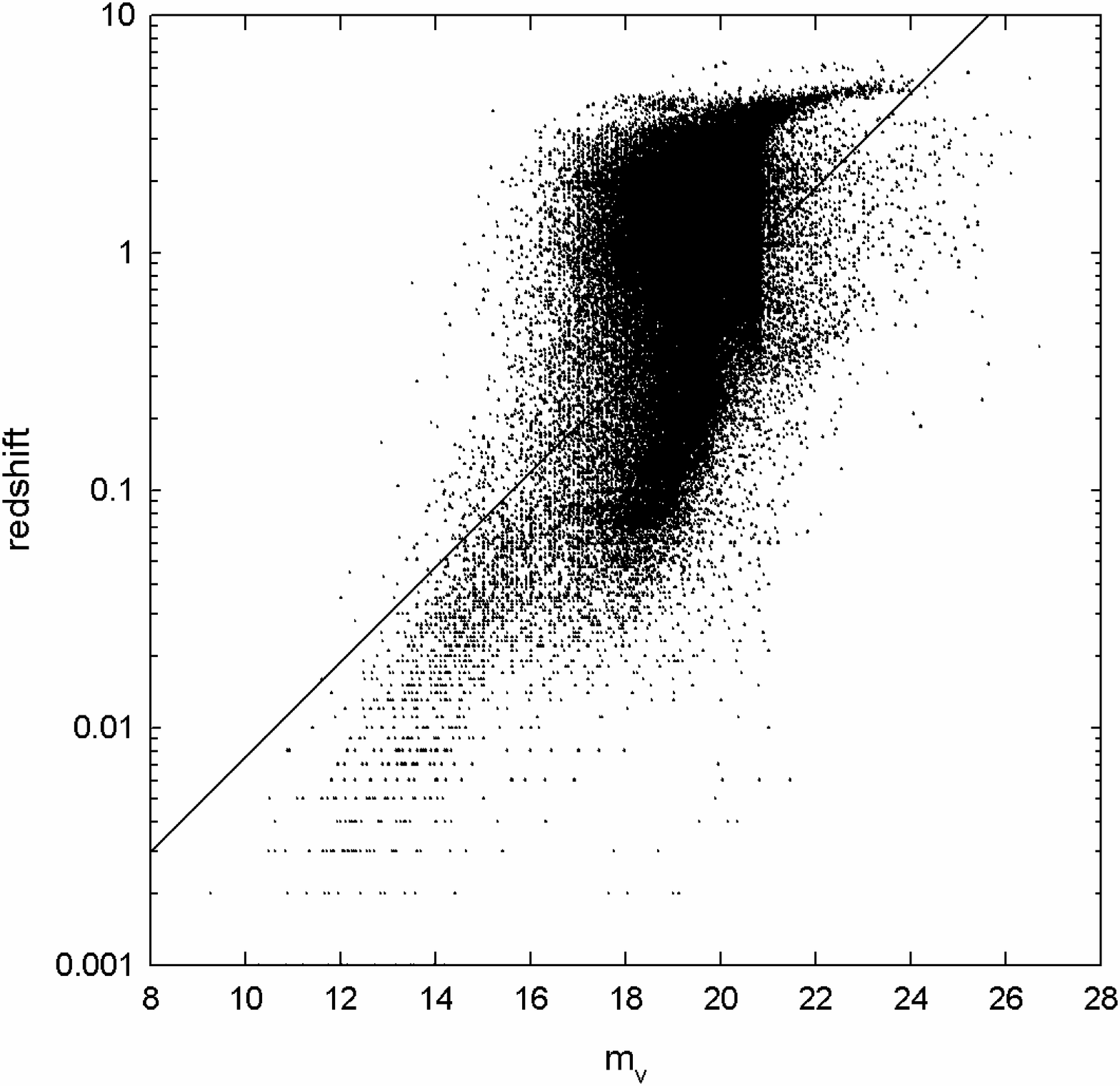}
\caption{A plot by M. Bell (2007) of 106,958 quasars and active
galaxies in the VCV Catalog. The solid line indicates Hubble relation
for first ranked cluster galaxies.  
\label{fig2}}
\end{figure}

\section{Evolution in the Hubble Diagram}

The standard Hubble line in Fig. 1 is delineated by E galaxies which
are the brightest in their clusters. The objects which fall above this 
line can be loosely referred to as ``active''. We will use in the
following discussion the fact that their underlying spectra comprise
non thermal radiation which cuts off further to the red as they
age. Observationally they span the extremes between the strong
synchrotron continuum of the quasars to the far red cut off of the
radio galaxies and finally to the oldest E galaxies which define the
accepted Hubble line in Figs.1 and 2. 

At this point we invoke the analyses of individual galaxy - quasar
associations which establish that the quasars start out their life at very
small luminosity and very high redshift (Arp 1998a, 2003). The general
solution of the field equations then show that the particles near zero
mass in recently created proto quasars increase as $t^2$ (t being the
age of the created particles in the variable mass hypothesis; Narlikar 1977;
Narlikar and Das 1980). In order to conserve momentum the particles
slow down, lowering the temperature and helping the growing gravity to
condense the plasma into a coherent proto galaxy (Actually the effect
for which ``dark matter'' was invented.) After this rapid wringing out
of the synchrotron energy, low mass atoms form and, because of
increasing mass of the electrons in energy level transitions, the
intrinsic wavelengths of emissiion and absorption, the redshifts,
decrease. (Narlikar and Arp 1993.) 

The resultant evolution in the z - m diagram is to brighten apparent
magnitude initially followed by a deep decline of intrinsic redshift.
Then a slower brightening of the object as star formation sets in.
Looking at the distribution of objects in Fig. 2 would suggest a fast
evolution in luminosity from about z = 6 to 4, then a long decline in
intrinsic redshift from z= 5 to 0.1 and finally a slow approach to the
Hubble line for high mass AGNs and somewhat below the Hubble line for
low mass progenitors. 

As they near the end point of their evolution (when their intrinsic
redshift has stabilized near zero) they have ended up somewhat under
the standard Hubble line (because most of them finish as intermediate
bright galaxies). If one were to shut off the galaxy creating
mechanism one would see the active galaxies disappearing with time
leaving only the standard Hubble line.

The key to this is the solution of the generalized field equations based
on the Hoyle Narlikar Machian gravitation theory as described by
Narlikar (1977). For a constant mass approximation the theory reduces to the
standard Einstein theory. However, the input from Mach's principle,
suggests that the inertia of a newly created particle starts off as zero
and grows with its age as it begins to get Machian contributions from more
and more remote matter in the universe. The typical wavelengths emitted by
a particle (such as the electron in a hydrogen atom) would reduce as its
mass grows. Thus newly created matter would exhibit high intrinsic
redshift. In such a framework, the intrinsic redshift of all galaxies
created at the same time as our own will always give a perfect Hubble
line because the look back time to a distant galaxy will always reveal
it at a younger age when its intrinsic redshift was exactly that
predicted by the Hubble law, cz = d x Ho.

In other words the Hubble line is the the line of repose, the
evolutionary end for all galaxies of the age of our own galaxy.
Younger galaxies have higher intrinsic redshifts which do not signal
their velocity or distance but only their age and their luminosity at
that time  which can be much fainter than our own, contrary to the
tenants of current astronomy.

\section{Definition of active galaxy and QSO/AGN}

The observational properties of the the quasars are their spectral
characteristics and their mainly stellar images. But in my
opinion an error in taxonomy was made when the term QSO was
arbitrarily {\it defined} as an object more luminous than, $M_V\leq$
-23 mag. (For example, as listed in the Cetty-V\'eron and V\'eron
Catalogue). 

This meant that the definition depended on a theory about redshifts,
not an observational property. The Nobelist Percy Bridgeman (1936)
stressed the necessity for science to use operational definitions. If
we follow that principle here we would suggest a definition along the
following lines:

{\it A QSO/AGN is A high surface brightness object with a non thermal energy
spectrum.} 

Surface brightness is a quantity which does not depend on distance so
we are not in the embarassing position of calling an object stellar at
low resolution then a galaxy at higher resolution. As to the second
criterion: If an object is composed of stars the spectral continuum
falls off in the blue and red as bodies do when radiating at their
effective temperature. If the energy spectrum is flatter the object is
not dominated by stars. The flat continua in QSOs have long been
identified as radiation from accelerating or decelerating charged
particles (e.g. synchrotron radiation).  Therefore we can empirically
classify an object as active from its energy spectrum alone. Also, of
course, by the presence of strong, broad, emission lines.

What to call the objects which look like QSOs but have moderate to low
redshift? As Fig. 1 shows they go by a variety of names but usage
required some blanket term and active galaxy nuclei (AGN) came to be
used for galaxies with compact energetic nuclei and AGN galaxy for
objects where the whole compact body appeared active (including
therefore QSOs). 

Fig. 3 shows these active QSO/AGN galaxies in the redshift - color
plot. It emphasizes the fact that regardless of whether non thermal or
thermal radiation is dominant, the evolution from QSO towards lower z
active galaxies is accompanied by evolution to redder colors right up
to the older galaxies of the kind which define the Hubble line.

\begin{figure}[ht]
\includegraphics[width=11.0cm]{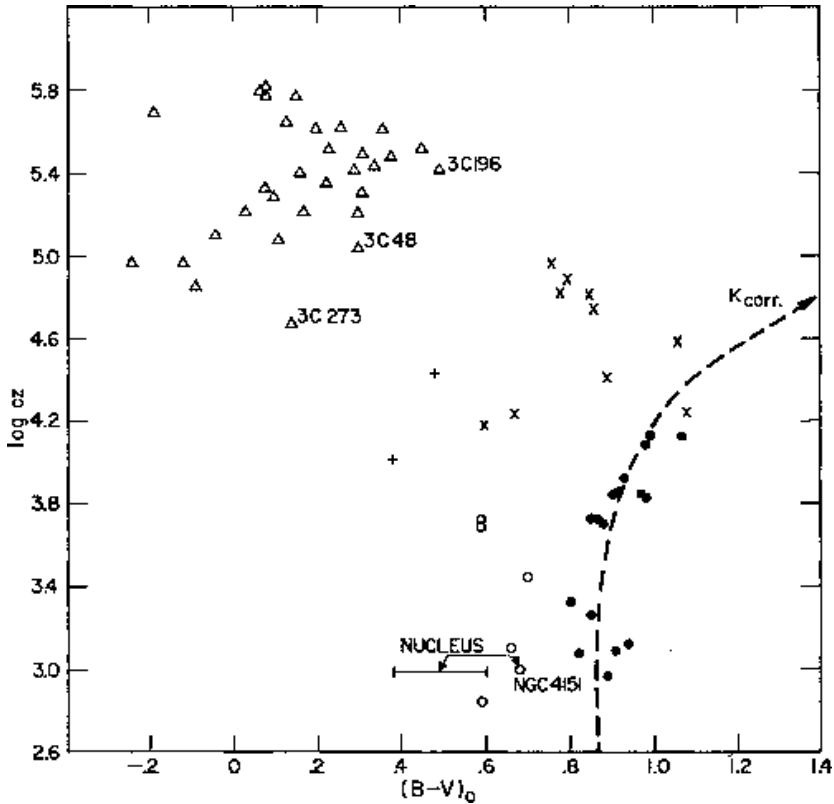}
\caption{Color evolution in the redshift - color plane. Triangles are
QSOs, crosses are radio galaxies with compact nuclei, two plus signs
are compact galaxies with Seyfert spectra, open circles represent
Seyfert galaxies and filled circles represent old E galaxies.
\label{fig3}}
\end{figure}

\subsection{Evolution of thermal and  non thermal radiation}

The defining characteristics of active galaxies can vary in relative
amount. For example, if we utilize the fact that high energy particle
radiation will decay faster than the low energy radiation we can
arrange QSOs in order of the age of their energy burst by their blue
cut off.  The high energy gamma and X-ray continua will quickly age
into ultraviolet excess QSOs which in turn will age into infra red
excess and finally into the longest lived, radio radiation. It is
strongly suggested here that QSOs are evolving proto galaxies. They pass
from a higher energy state of an ionized gas to beginning formation of
atoms - from optical entities called Quasi Stellar Objects through
compact radio galaxies and finally to normal, quiescent appearing galaxies. 

How does this fit with the observations? A very few QSO's are strong
gamma ray sources, more are X-ray sources, then UV excess QSO's and
finally radio QSO's (originally called QSR's). Emphasis on the non
thermal energy source enables an empirical model of the QSO phenomenon
from the most energetic to the quietest extra galactic objects. It is
interesting that this picture looks much like the conventional one of
a QSO as the nucleus of a host galaxy. And indeed the nucleus of an
active galaxy (AGN) would fit a QSO as we have defined it. If smaller
in luminosity, however, a QSO could also be considered a smaller
portion of an active nucleus with an initially higher redshift

\subsection{Evolution of stellar content}

Once star production has started in a QSO/AGN galaxy the spectrum
can come to be dominated by thermal radiation. In such cases, however,
the blueness of the optical spectrum still reflects bright, hot stars
and a stellar population which is younger than, say a quiescent
galaxy. Evolved red giant stars can be conversely used as an
indicator of an old stellar population. This evolutionary continuity
is well illustrated here in Fig.4. 

\begin{figure}[ht]
\includegraphics[width=10.0cm]{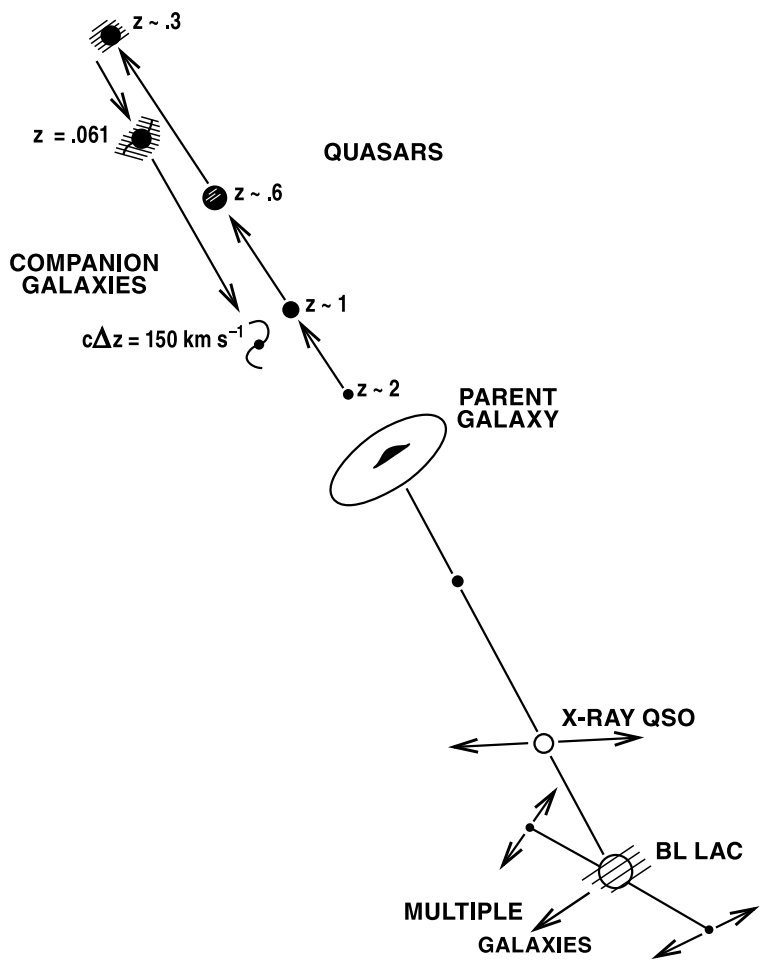}
\caption{Summary of empirical data on evolution from high redshift
quasar to low redshift companion galaxy. From Arp (1998b).   
\label{fig4}}
\end{figure}

\subsection{Luminosity of QSOs}

In our operational definition of QSO we have not included the property
of luminosity because the meaning of redshift has not been agreed
on. QSO's do not obey a redshift-apparent magnitude relation. In order
to obey a redshift-distance relation their luminosity range would have
to be enormous. At present therefore the only way we could estimate
distances to QSOs is by association with other extragalactic objects
whose distances we accept.

Associations of high redshift quasars with low redshift galaxies have
been accumulating for more than 40 years. But the majority of
astronomers still firmly believe that QSO redshifts measure greater
distances and hence have much greater luminosities. Instead of
redshifts caused by recessional velocities in an expanding universe,
however, we try to review here the evidence that they are intrinsic
redshifts caused by younger matter in the QSOs. 

In spite of General Relativity's assumption that matter in the
Universe is homogeneously distributed, the galaxies are conspicuously
grouped on the sky in clusters and super clusters. QSOs are more
widely spread but also show clustering (although the somewhat similar
redshifts differ too much to be velocity.) Denied for many years,
current measures show a correlation with QSO's on the scale of about 1
degree. This is now interpreted as gravitational lensing of background
QSOs by foreground galaxies. In a discussion of their own and 23 other
analyses J. Nollenberg and L. Williams (2005) appear to accept
correlations of galaxies and higher redshift quasars. But in
commenting on possible gravitational lensing effects they note that
greater amounts of cold dark matter are needed than in currrent
models. Of course the alignments, pairings and connections of
companion quasars to much lower redshift galaxies which
are observed would exclude lensing. (Arp and Crane 1992; Arp 1998a,
p177)

In the real world, however, galaxies are almost never isolated. The
typical group consists of a dominant, lowest redshift galaxy with
increasingly fainter and higher redshift companions up to and
including QSOs. More and brighter QSO's are found around galaxies 
which have active (strong, non thermal) nuclei. The question is what
is different about the companions compared to the dominant galaxy?
The answer is the smaller galaxies tend to have younger stars, are less
dynamically relaxed and more active. What more obvious conclusion
could be made other than that they are younger? Add to this the fact
that Erik Holmberg found as long ago as 1969 that companion galaxies
tended to lie along the minor axes of disk galaxies. Is there any
alternative to the their having originated in the nuclei of these
larger galaxies and  escaped along the line of least resistance? The
evolution of quasars into companion galaxies was then strongly
supported when it was shown that: ``Pairs of quasars tend to lie even
more closely along minor axis of ejecting galaxies.''(Arp 1998a) The
quasar result was confirmed by L\'opez-Corredoira and Guti\'errez
(2006) with a larger sample of quasars. 

\begin{figure}[ht]
\includegraphics[width=11.0cm]{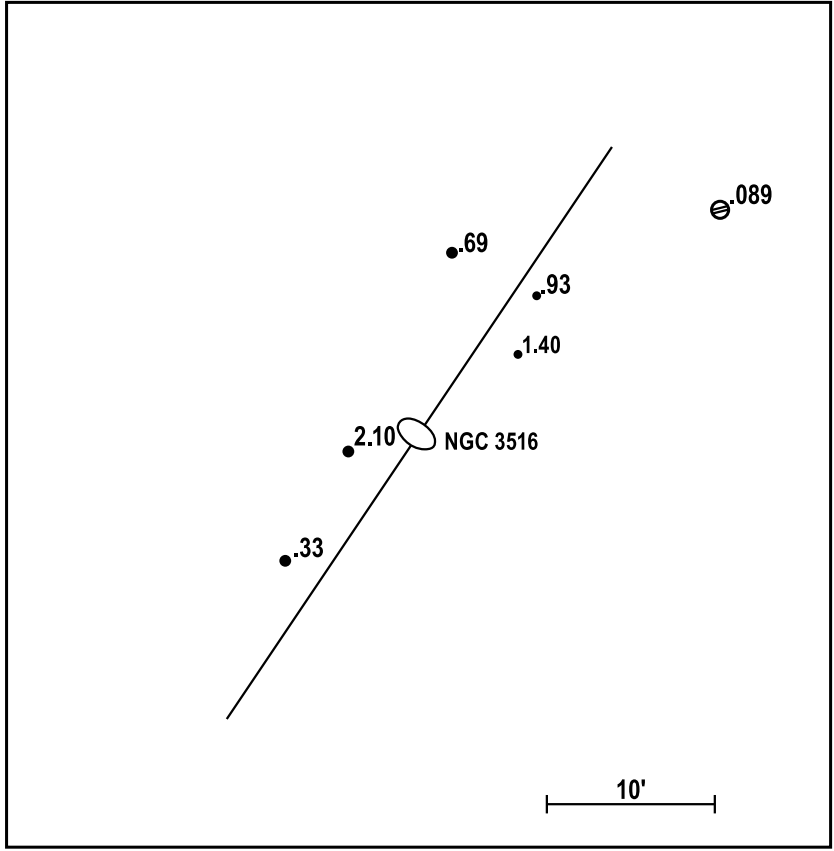}
\caption{Rosetta stone of ejected quasars. The brightest X-ray sources
in the field are quasars coming out along the minor axis of the
Seyfert which is NGC 3516 at z = .009 
\label{fig5}}
\end{figure}

 One of the most striking cases is shown here in Fig. 5.
The Chinese astronomers Chu and Xu realized the position of these six
strong X-ray sources warranted their taking spectra of each optical 
identification. The closeness brightness and alignments identified
them clearly as belonging to the central active galaxy. They turned
out to be have apparent magnitudes from 18.5 to 20.2 mag. and if NGC
3516 has an absolute magnitude of $M_V$ = -20.5 the five quasars would
have luminosities of $M_V$ -13 to -14. So the absolute magnitudes of the
quasars are not in the range above -23 as they are supposed by current
definition to be but instead are only three or four magnitudes
brighter than the brightest stars in a low redshift spiral galaxy.
The brighter, outer, X-ray compact galaxy (z = .089) is $M_V$ = -18.2
which empirically confirms the evolution from quasar into active
companion galaxy. Because of the commonly accepted $M_V$ -23mag. lower
limit in the definition of a quasar, however, it makes it very
difficult to now relinquish the picture of a QSO as a super luminous
galaxy and switch to the observational evidence for it being a small
seed with its redshift as a parameter of evolution.

{\bf Summary}

Quasars and active galaxies form a broad, continuous track in the
Hubble diagram. It is argued that the evolution of QSOs must be along
this sequence in both intrinsic redshift and apparent magnitude. If
so, the evolution comes to rest near the age of our galaxy on the
conventional Hubble redshift - apparent magnitude relation. It is
considered highly significant that the more general solution of
the Einstein field equations by Narlikar predicts evolution in
intrinsic redshift and also exactly the Hubble relation for all galaxies
created at the same time as our own galaxy.

{\bf References}

Arp, H. 1968, ApJ 153, L33

Arp, H. 1998a, Seeing Red, Apeiron, Montreal

Arp, H. 1998b, ApJ 496, 661

Arp, H. 2003, Catalogue of Discordant Redshift Associations, Apeiron, Montreal

Arp, H. and Crane, P. 1992, Phys. Lett.A, 168, 6 

Bell, M. 2007, ApJ 667, L129

Bridgeman, P. 1936, The Nature of Physical Theory'', John Wiley and Sons

Chu, Y., Wei, J., Hu J., Zhu, X., Arp, H. 1998, ApJ 500, 596

L\'opez-Corredoira, M. and Guti\'errez. C. 2007, A\&A 461, 59 

Narlikar J. 1977, Ann. Phys. 107, 325

Narlikar J. and Das, P. 1980, ApJ 240, 401

Narlikar J., Arp H. 1993 ApJ 405, 51

Nollenberg, J., and Williams, L. 2005, ApJ 634, 793

\end{document}